\documentclass[12pt]{article}

\usepackage[utf8x]{inputenc}
\usepackage[english]{babel}

\usepackage{amsmath}
\usepackage{amssymb}
\usepackage{graphicx}
\usepackage{multicol}
\usepackage{url}
\usepackage{caption}
\usepackage{subcaption}

\RequirePackage[colorlinks,citecolor=blue,urlcolor=blue,linkcolor=blue]{hyperref}
\usepackage{doi} 

\newcommand\ee{e^+e^-}

\begin{document}

\title{ \bf Fully Geant4 compatible package for the simulation of Dark Matter in fixed target experiments}

\author{M.~Bondi$^{1}$, A.~Celentano$^{1}$\footnote{{\bf e-mail}: andrea.celentano@ge.infn.it},
     R.~R.~Dusaev$^{2}$\footnote{{\bf e-mail}: renat.dusaev@cern.ch},
    D.~V.~Kirpichnikov$^{3}$\footnote{{\bf e-mail}: dmbrick@gmail.com}, \\
    M.~M.~Kirsanov$^{3}$\footnote{{\bf e-mail}: mikhail.kirsanov@cern.ch},
     N.~V.~Krasnikov$^{3,4}$\footnote{{\bf e-mail}: nikolai.krasnikov@cern.ch}, L.~Marsicano$^{1}$, D. Shchukin$^5$
\\
 $^1$ Istituto Nazionale di Fisica Nucleare, Sezione di Genova, 16146 Genova, Italy\\
 $^2$ Tomsk Polytechnic University, 634050 Tomsk, Russia \\
 $^3$ Institute for Nuclear Research of the Russian Academy of Sciences, \\117312 Moscow, Russia \\
 $^4$ Joint Institute for Nuclear Research, 141980 Dubna, Russia \\
  $^5$ P.N. Lebedev Physical Institute, Moscow, Russia, 119991 Moscow, Russia \\
 }

\date{\today}

\maketitle

\begin{abstract}
 We present the package for the simulation of DM (Dark Matter) particles in fixed target experiments. The most convenient way
of this simulation (and the only possible way in the case of beam-dump) is to simulate it in the framework of the
Monte-Carlo program performing the particle tracing in the experimental setup.
The Geant4 toolkit framework was chosen as the most popular and versatile solution nowadays.

 Specifically, the package includes the codes for the simulation of the processes of DM particles production via electron and muon bremsstrahlung
off nuclei, resonant in-flight positron annihilation on atomic electrons and gamma to ALP (axion-like particles) conversion on nuclei.
Four types of DM mediator particles are considered: vector, scalar, pseudoscalar and axial vector.
The total cross sections of bremsstrahlung processes are calculated numerically at exact tree level (ETL).

 The code handles both the case of invisible DM decay and of visible decay into $e^+e^-$ ($\mu^+\mu^-$ for $Z'$, $\gamma \gamma$ for ALP).

 The proposed extension implements native Geant4 application programming interfaces (API) designed for these needs and can be unobtrusively
embedded into the existing applications.
 

 As an example of its usage, we discuss the results obtained from the simulation of a typical active beam-dump experiment.
We consider $5 \times 10^{12}$ 100 GeV electrons impinging on a lead/plastic heterogeneous calorimeter playing a role of an active
thick target. The expected sensitivity of the experiment to the four types of DM mediator particles mentioned above is then derived.

\end{abstract}

\section*{Program summary}

\emph{Program title:} DMG4 \\
\emph{CPC Library link to program files:} \\
\emph{Code Ocean capsule:} \\
\emph{Licensing provisions:} GNU General Public License 3 (GPL) \\
\emph{Programming language:} c++ \\
\emph{Nature of problem:} The optimal way to simulate Dark Matter production processes in fixed target experiments in most cases is to do it
 inside the program for the full simulation of the experimental setup and not separately, in event generators. The code that can be easily
 embedded in such programs is needed. The code should be able to simulate various DM production processes that happen in a thick target,
 in particular on nuclei, with maximal accuracy. \\
\emph{Solution method:} We created a Geant4 compatible DM simulation package for this purpose. The choice of this simulation framework
 is suggested by its popularity and varsatility. The code includes the cross sections precalculated at exact tree level for a wide variety
 of DM particles.

\section{Introduction}

Models with light Dark Matter (DM) particles are very popular in the searches for physics beyond the Standard Model (SM).
The light dark matter (LDM) hypothesis conjectures the existence of a new class of lighter elementary particles, not charged under the SM interactions.
The simplest model predicts LDM particles (denoted as $\chi$) with masses below 1~GeV/c$^2$, charged under a new force in Nature and interacting with
the SM particles via the exchange of a light mediator. In the simplest model, the mediator is a $1^-$ vector boson, usually referred to
as ``heavy photon'' or ``dark photon''~\cite{HOLDOM1986196}. However, relevant model variations correspond to different mediator quantum
number assignments. This picture thus foresees the existence of a new ``Dark Sector'' in Nature, with its own particles and interactions,
and is compatible with the well-motivated hypothesis of DM thermal origin. A complete introduction to this subject can be found,
for example, in the 2017 US Cosmic Visions community report~\cite{mb}, or in the 2019 CERN Physics Beyond Colliders report~\cite{Beacham:2019nyx}. 

Accelerator-based thick-target experiments at moderate beam energy ($\sim$ 10$\div$100 GeV) are the ideal tool to probe the new hypothesis since
they have a very large discovery potential in a wide area of parameters space. On the other hand, direct detection efforts typically show
a limited sensitivity to LDM due to the very low energy of the recoil, often lower than the detection threshold. 

In many cases such searches are performed in (active) beam-dump experiments~\cite{jdb,PhysRevD.88.114015,PhysRevD.80.095024,PhysRevD.91.094026}.
In these experiments, many different processes can result in DM production inside the thick target with initial particles at a wide spectrum
of energies and topologies, due to the production of secondaries from the primary impinging particle. 
Therefore, the optimal way to simulate these processes is to do it inside the program for the full simulation of the experimental setup,
to account for the correlation among the initial-state particles kinematic variables and to fully take into account the production
cross-section dependence on these.

We created a Geant4 compatible package for the simulation of various types of DM production -- the choice of this simulation framework
was suggested by the fact that, today, it is the most versatile and mature popular toolkit for full simulation programs
used in HEP experiments~\cite{geant4} designed to maintain full lifecycle of HEP experiments. The package is named \emph{DMG4}.
The code tends to follow the Geant4 API conventions as close as possible.

\section{DMG4 package structure}

\begin{figure*}[h]
\begin{center}
\includegraphics[width=0.75\textwidth]{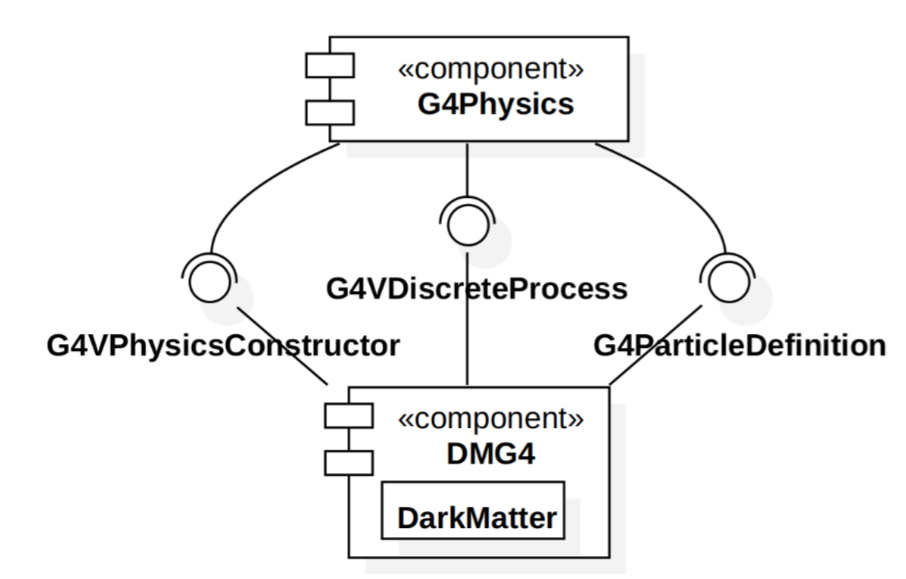}
\caption {Component diargam of the DMG4 package.
\label{fig:Diagram1}}
\end{center}
\end{figure*}

The DMG4 package is a cohesive set of DM particle definition classes, DM process classes and the DM physics class that
assembles all together. Historically, it includes a separate package \emph{DarkMatter} with a collecton of cross section calculation
routines. This package was used previously through the Geant4 classes inherited from \texttt{G4UserSteppingAction} and
\texttt{G4UserRunAction}. The package structure is illustrated in Figure~\ref{fig:Diagram1}.
The new particles introduced so far in the package are listed in Table~\ref{table:dmparticles}.
The PDG codes are ascribed according to the slightly extended rules in~\cite{PDG_particles}.

\begin{table}[h]
\caption{DM particles defined in the package DMG4}
\begin{center}
\begin{tabular}{|c|c|c|c|c|c|c|}
\hline
Name                      &  PDG ID      & emitted by & spin & parity & stable? & decay           \\
\hline
DMParticleAPrime          &  5500022     & $e^+,e^-$  &  1   &  1     & true    & -               \\
\hline
DMParticleXBoson          &  5500122     & $e^+,e^-$  &  1   &  1     & false   & $e^+e^-$        \\
\hline
DMParticleScalar          &  5400022     & $e^+,e^-$  &  0   &  1     & true    & -               \\
\hline
DMParticleXScalar         &  5400122     & $e^+,e^-$  &  0   &  1     & false   & $e^+e^-$        \\
\hline
DMParticlePseudoScalar    &  5410022     & $e^+,e^-$  &  0   & -1     & true    & -               \\
\hline
DMParticleXPseudoScalar   &  5410122     & $e^+,e^-$  &  0   & -1     & false   & $e^+e^-$        \\
\hline
DMParticleAxial           &  5510022     & $e^+,e^-$  &  1   & -1     & true    & -               \\
\hline
DMParticleZPrime          &  5500023     & $\mu$      &  1   &  1     & true    & -               \\
\hline
DMParticleALP             &  5300122     & $\gamma$   &  0   & -1     & false   & $\gamma \gamma$ \\
\hline
\end{tabular}
\end{center}
\label{table:dmparticles}
\end{table}

The dark sector particles that are used for the missing energy signature simulations are assumed to be stable, although
in full models they could decay into other dark matter particles. 
However, this is unimportant as long as they are also invisible and carry away energy -  for this reason, in the following we will
call generically ``dark matter'' the dark sector particles produced in the detector. The extension to the case of partly visible
DM decay products that could be observed through cascade decays is straightforward.

The current version of DMG4 package contains the following processes of DM production:

\begin{itemize}
\item
 Bremsstrahlung-like process of the type $b N \to b N X$, where $b$ is a projectile (can be $e^-, e^+, \mu^-, \mu^+$),
and $X$ is a DM particle

\item
 Primakoff process of photon conversion $\gamma N \to a N$, where $a$ is an axion-like particle (ALP)~\cite{dusaev2020photoproduction}

\item
 Resonant in-flight positron annihilation on atomic electrons $e^+ e^- \rightarrow X \rightarrow \chi \chi$, where $\chi$ is a dark
matter mediator decay product~\cite{PhysRevLett.121.041802}.

\end{itemize}

 In the latter case, the DM particle $X$ acts as a $s-$channel intermediate resonance, with a non-zero intrinsic width due to the decay
to final state invisible particles. For missing energy signature simulations, as discussed before, the role of the decay
products $\chi$ is the same as the role of the DM particle $X$ in the previous production mechanisms, since they carry away energy from
the active target without being detected.

 The physics for a simulation run is configured in the function \texttt{DarkMatterPhysicsConfigure} called from the constructor
of the factory class \texttt{DarkMatterPhysics}. One has to create an instance of one of the concrete classes corresponding to the needed
process and derived from the base class \texttt{DarkMatter}, for example \texttt{DarkPhotons}.
The factory then instantiates and registers the needed particles and processes provided by the DMG4 package in terms of the native Geant4 API.
The required parameters include the mixing parameter $\epsilon$ and cut-off minimal energy of particles that can initiate the processes of
DM production. The latter is needed to avoid simulation of very soft DM particles that are anyway undetectable.

 As in many other Geant4 physics classes, there is a parameter that can bias the production cross section, i.e. increase it in such
a way that the fraction of events with DM production is not too small. The simulation without biasing is practically impossible
as for physically interesting values of $\epsilon$ one would have to simulate too many events to have sufficient statistics.
At the same time the fraction of events with DM production should be significantly smaller than 1, otherwise the energy and coordinate
distributions can be distorted. It is recommended in any case to keep it smaller than 0.07, for some processes smaller than 0.03.

 The \emph{DarkMatter} package contains the routines that calculate the cross sections, total and differential. This is explained in more details
in the next section.

\section{Package DarkMatter and ETL cross sections}

The formulas for the cross sections, total and differential, implemented in the package are derived for different cases.
For the bremsstrahlung-like and the $\ee$ annihilation processes we consider the following scenarios, with different
quantum number assignments for the DM mediator particles \cite{Liu_2017,Liu_2017_2}, assuming for simplicity
that all other DM particles $\chi$, coupled only to these mediators, are fermions.
\\
Vector case: 
\begin{equation}
\mathcal{L} \supset \mathcal{L}_{SM} -\frac{1}{4} V_{\mu\nu}^2 +\frac{1}{2} m_V^2 V_\mu^2 +
 \sum_{\psi} e \epsilon_V V_\mu  \bar{\psi} \gamma^\mu \psi  + 
 g^D_V V_\mu \bar{\chi}  \gamma^\mu \chi + \bar{\chi}(i\gamma^\mu \partial_\mu - m_\chi)\chi
 \label{VecLagr1}
\end{equation}
Axial vector case:
\begin{equation}
\mathcal{L} \supset \mathcal{L}_{SM} -\frac{1}{4} A_{\mu\nu}^2 +\frac{1}{2} m_A^2 A_\mu^2 +
 \sum_{\psi} e \epsilon_A A_\mu  \bar{\psi} \gamma_5 \gamma^\mu \psi  + 
 g^D_A  A_\mu \bar{\chi} \gamma_5 \gamma^\mu \chi 
 + \bar{\chi}(i\gamma^\mu \partial_\mu - m_\chi)\chi
 \label{AxialVecLagr1}
\end{equation}
Scalar case:
\begin{equation}
\mathcal{L} \supset \mathcal{L}_{SM} + \frac{1}{2} (\partial_\mu S)^2 - \frac{1}{2} m_S^2 S^2 +
 \sum_{\psi} e \epsilon_S S  \bar{\psi} \psi  + 
 g^D_S  S \bar{\chi}  \chi 
 + \bar{\chi}(i\gamma^\mu \partial_\mu - m_\chi)\chi
 \label{ScalLagr1}
\end{equation}
Pseudo-scalar case:
\begin{equation}
\mathcal{L} \supset \mathcal{L}_{SM} + \frac{1}{2} (\partial_\mu P)^2 - \frac{1}{2} m_P^2 P^2 +
 \sum_{\psi} i e \epsilon_P P  \bar{\psi} \gamma_5 \psi  + 
 g^D_P  P \bar{\chi} \gamma_5 \chi 
 + \bar{\chi}(i\gamma^\mu \partial_\mu - m_\chi)\chi,
 \label{PseudoscalLagr1}
\end{equation}
where $\epsilon_V,\epsilon_A,\epsilon_S,\epsilon_P$ are the mixing (or coupling) parameters, $m_V,m_A,m_S,m_P$ are the masses of mediators.
\\
For ALPs, instead, we consider the simplified model \cite{Dobrich_2016} with ALP coupling predominantly to photons:
\begin{equation}
\mathcal{L}_{int} \supset - \frac{1}{4} g_{a \gamma \gamma} a F_{\mu \nu} \tilde{F}^{\mu \nu} +
\frac{1}{2}(\partial_\mu a)^2-\frac{1}{2} m_a^2 a^2,
\label{ALPlagr1}
\end{equation}
where $F_{\mu \nu}$ denotes the strength of the photon field, and the dual tensor is defined by
$\tilde{F}_{\mu \nu}  = \frac{1}{2} \epsilon_{\mu \nu \lambda \rho} F^{\lambda \rho}$.
We assume that the effective coupling, $g_{a \gamma \gamma}$, and the ALP mass, $m_a$,
are independent.

For the electron bremsstrahlung process, the simulation package contains the analytical expressions for the cross sections,
total and differential, derived in the IWW (improved Weizsaker-Williams) approximation \cite{jdb}.
However, as discussed already in \cite{DMsimulation}, these can be rather inexact in some regions of parameter space.
For this reason, the package contains the tabulated K-factors that correct the total cross sections to the
values calculated in ETL (exact tree-level) limit \cite{Liu_2017,Liu_2017_2,DMsimulation}. The total ETL cross-sections
were pre-calculated using the means of symbolic computation software Mathematica~\cite{Mathematica}. As compared to \cite{DMsimulation},
we extended the tables with K-factors to the cases of scalar, pseudoscalar and axial vector DM mediator particles. At runtime, the total
cross is obtained from the tabulated values using the interpolation. The differential cross section formulas are shown in Appendix A.
The tabulated differential cross sections are also used in some limited regions, where the difference is significant.

For the $\ee$ annihilation process the following expression for the production cross section is implemented in the code:

\begin{equation}
\sigma_{\ee}=\frac{4\pi \alpha_{EM} \alpha_D \varepsilon^2}{\sqrt{s}}q\frac{\mathcal{K}}{(s-m_X^2)^2+\Gamma_X^2m_X^2} \; \;
\end{equation}
where $s$ is the invariant mass of the $\ee$ system, $m_X$ the mass of the intermediate DM particle
(where $X=V,A,S,P$), $q=\frac{\sqrt{s}}{2}\sqrt{1-\frac{4m_\chi^2}{s}}$, $\Gamma_X$ is the intermediate DM particle
decay width to dark particles $\chi$, discussed in the following, $\alpha_{EM}$ is the electromagnetic fine structure constant,
and $\alpha_D\equiv \frac{\left(g^D_X\right)^2}{4\pi}$ is the coupling squared to the dark particles $\chi$.
Finally, $\mathcal{K}$ is a kinematic factor that reads, respectively, $(s-\frac{4}{3}q^2)$ for the vector DM,
$\frac{8}{3}q^2$ for the axial vector case, $2q^2$ for the scalar case, and $\frac{s}{2}$ for the pseudo-scalar case.
These expressions correspond to the exact tree-level calculation, with the replacement
$(s-m^2_{X})^2 \rightarrow (s-m^2_{X})^2+\Gamma_X^2m^2_{X}$ in the last denominator to regulate the tree-level
cross-section divergence at the resonance pole.

The following tree-level expressions for the decay widths are implemented. For the visible decay width,
valid for $m_{X}>2~m_e$, the vector, axial-vector, scalar, and pseudo-scalar cases read:
\begin{eqnarray}
\Gamma_{V\rightarrow e^-e^+} = \frac{\alpha_{QED}\epsilon^2}{3} m_{V}\bigl(1+
\frac{2m_e^2}{m_{V}^2}\bigr)\sqrt{1-\frac{4m_e^2}{m_{V}^2}},
\\
\Gamma_{A\rightarrow e^-e^+} = \frac{\alpha_{QED}\epsilon^2}{3} m_{A}\left(1-\frac{4m_e^2}{m_{A}^2}\right)^{3/2},
\\
\Gamma_{S\rightarrow e^-e^+} = \frac{\alpha_{QED}\epsilon_S^2}{2} m_{S}\left(1-\frac{4m_e^2}{m_{S}^2}\right)^{3/2},
\\
\Gamma_{P\rightarrow e^-e^+} = \frac{\alpha_{QED}\epsilon_P^2}{2} m_{P}\left(1-\frac{4m_e^2}{m_{P}^2}\right)^{1/2}
\end{eqnarray}

For the invisible decay width we have instead:\\
\begin{eqnarray}
\Gamma_{V\rightarrow \bar{\chi}\chi} = \frac{\alpha_D}{3} m_{V}\bigl(1+
\frac{2m_{\chi}^2}{m_{V}^2}\bigr)\sqrt{1-\frac{4m_{\chi}^2}{m_{V}^2}}, 
\\
\Gamma_{A\rightarrow \bar{\chi}\chi} = \frac{\alpha_D}{3} m_{A}\left(1-\frac{4m_{\chi}^2}{m_{A}^2}\right)^{3/2}, 
\\
\Gamma_{S\rightarrow \bar{\chi}\chi} = \frac{\alpha_{D}}{2} m_{S}\left(1-\frac{4m_\chi^2}{m_{S}^2}\right)^{3/2},
\\
\Gamma_{P\rightarrow \bar{\chi}\chi} = \frac{\alpha_{D}}{2} m_{P}\left(1-\frac{4m_\chi^2}{m_{P}^2}\right)^{1/2}.
\end{eqnarray}

The ALP coupled to photons (\ref{ALPlagr1}) has the following decay width
\begin{equation}
\Gamma_{a\rightarrow \gamma \gamma} = \frac{g_{a \gamma \gamma}^2 m_a^3}{64 \pi}.
\label{WidthALP}
\end{equation}

\section{Calculation of sensitivity of a typical active beam-dump experiment to various types of DM particles}

 We used the DMG4 package described above to calculate the sensitivity to various types of DM of a typical experiment
that uses a missing energy signature in the electron beam and compare them for the same beam energy
and EOT (number of electrons on target). We define the sensitivity as the expected 90\% C.L. upper limit on the parameter $\epsilon$
in the case of no signal and very small background. We perform the calculations for the typical energy of the electron beam
at the CERN SPS of 100 GeV and a lead/plastic electromagnetic calorimeter ECAL \cite{na64-prd} as an active target.

 As only one of the scenarios defined in Section 3 can be chosen for a single simulation run of the package, in the
following instead of $\epsilon_V,\epsilon_A,\epsilon_S,\epsilon_P$ we use simply $\epsilon$.

\begin{figure*}[h]
\begin{center}
\includegraphics[width=0.75\textwidth]{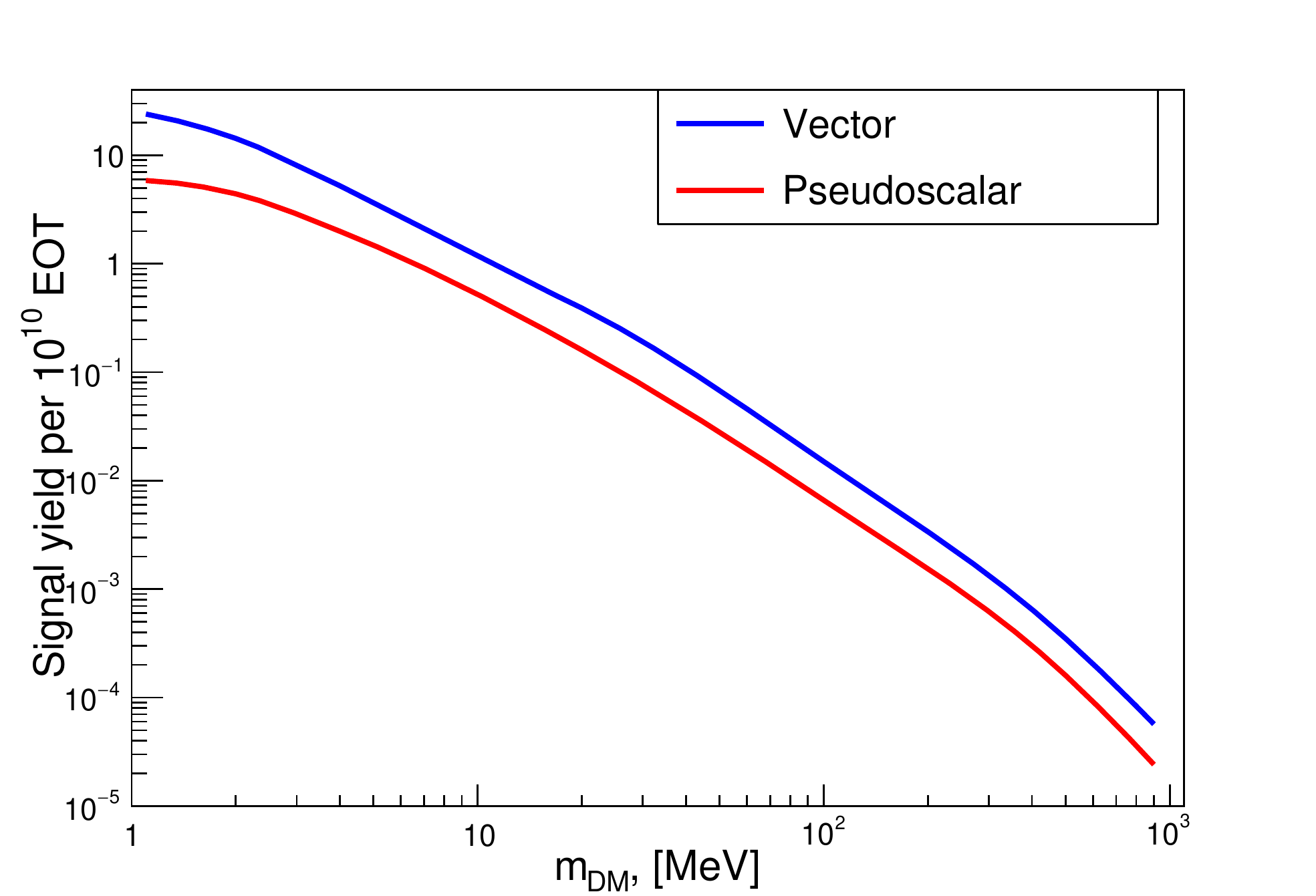}
\caption {The signal yield per $10^{10}$ EOT in the missing energy signature experiment for vector and pseudoscalar DM.
\label{fig:Yield}}
\end{center}
\end{figure*}

 In these estimations a signal event is an event with energy deposition in the ECAL smaller than 50 GeV and no
significant energy deposition (less than 1 GeV) in the hadron calorimeter installed downstream the ECAL. The number
of such signal events (signal yield in the following) produced in the electron beam for the mixing parameter $\epsilon = 10^{-4}$, calculated
for the vector DM (dark photon) and pseudoscalar DM according to cross sections from the package DarkMatter, is shown in
Table~\ref{table:tabCompSignWithCS} and Figure~\ref{fig:Yield}. In these calculations only bremsstrahlung
processes are taken into account. The difference between vector and pseudoscalar particles is significant.

\begin{table}[h]
    \caption{Comparison of the signal yields for the vector (VC) and pseudoscalar (PS) cases, per $10^{10}$ EOT.
             The cross section ratio calculated for the electron energy 100 GeV is also shown.}
    \begin{center}
    \begin{tabular}{|c|c|c|c|c|}
    \hline
$M_A$ [MeV] &  $N^{VC}_{sign}$  & $N^{PS}_{sign}$ & $N^{VC}_{sign}/N^{PS}_{sign}$ & $\sigma^{VC}_{tot}/\sigma^{PS}_{tot}$\\
    \hline
    1.1 &24.0    &5.85 &4.1 &4.12 \\
    \hline
    2	&14.3	&4.41 &3.2 &3.53 \\
    \hline
    4	&5.23	&1.99 &2.6 &3.114 \\
    \hline
    16.7 &0.516 &0.205 &2.51 &2.66  \\
    \hline
    20	&0.41	&0.16 &2.5 &2.64 \\
    \hline
    100	&0.015	&0.0066 &2.3 &2.47 \\
    \hline
    500	&0.00035 &0.00016 &2.2 &2.39 \\     
    \hline
    900	&0.00005685 &0.0000241 &2.36 &2.34 \\
    \hline
    \end{tabular}
    \end{center}
    \label{table:tabCompSignWithCS}
\end{table}

\begin{figure*}[h]
\begin{center}
\includegraphics[width=0.75\textwidth]{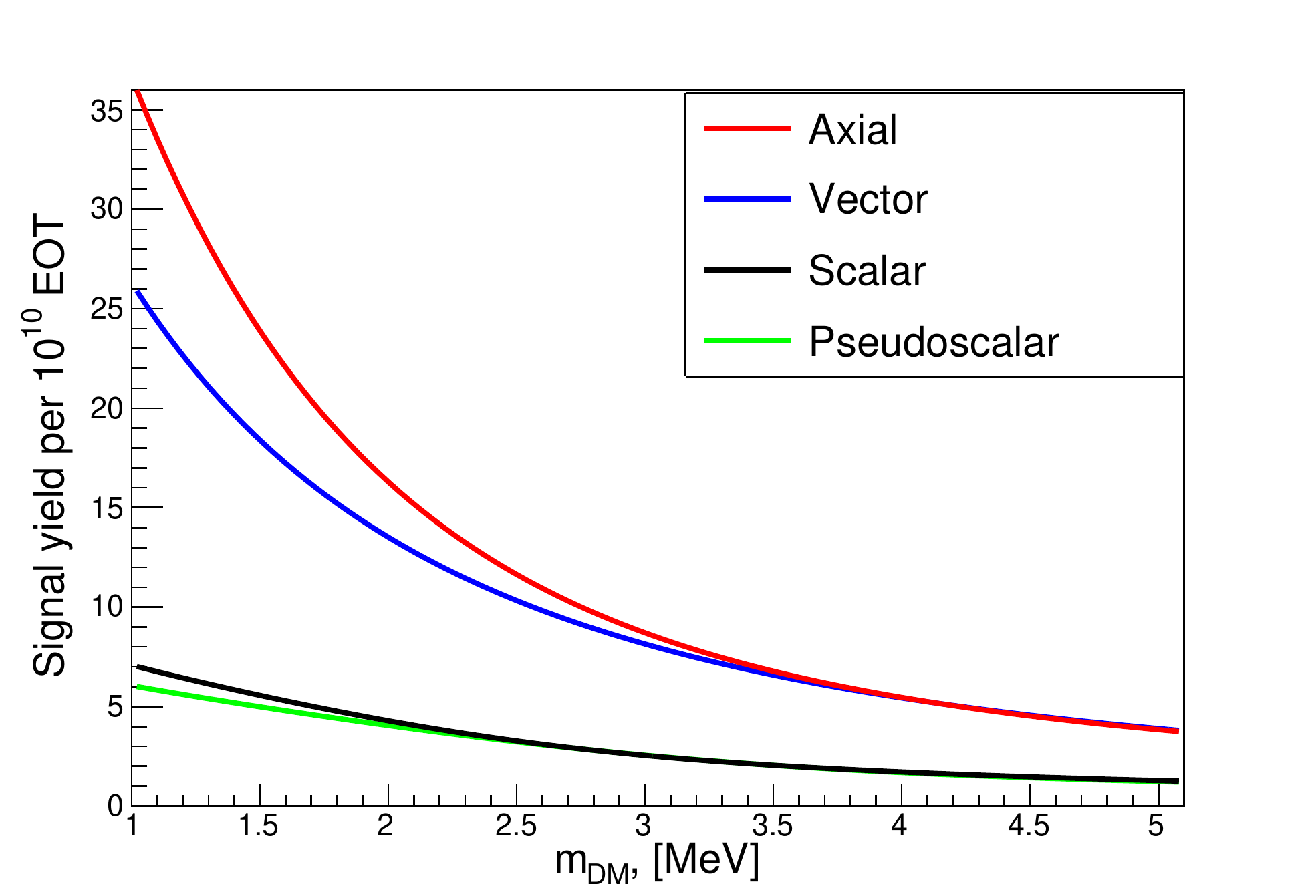}
\caption {The signal yield per $10^{10}$ EOT in the missing energy signature experiment for four types of DM.
\label{fig:Yield1}}
\end{center}
\end{figure*}

 The difference in the signal yield between vector and axial-vector DM is rather small; between scalar and pseudoscalar DM
it is still smaller. We show them separately in Figure~\ref{fig:Yield1}. The difference is significant only for the masses
below 4 MeV.

\begin{figure*}[h]
\begin{center}
\includegraphics[width=0.75\textwidth]{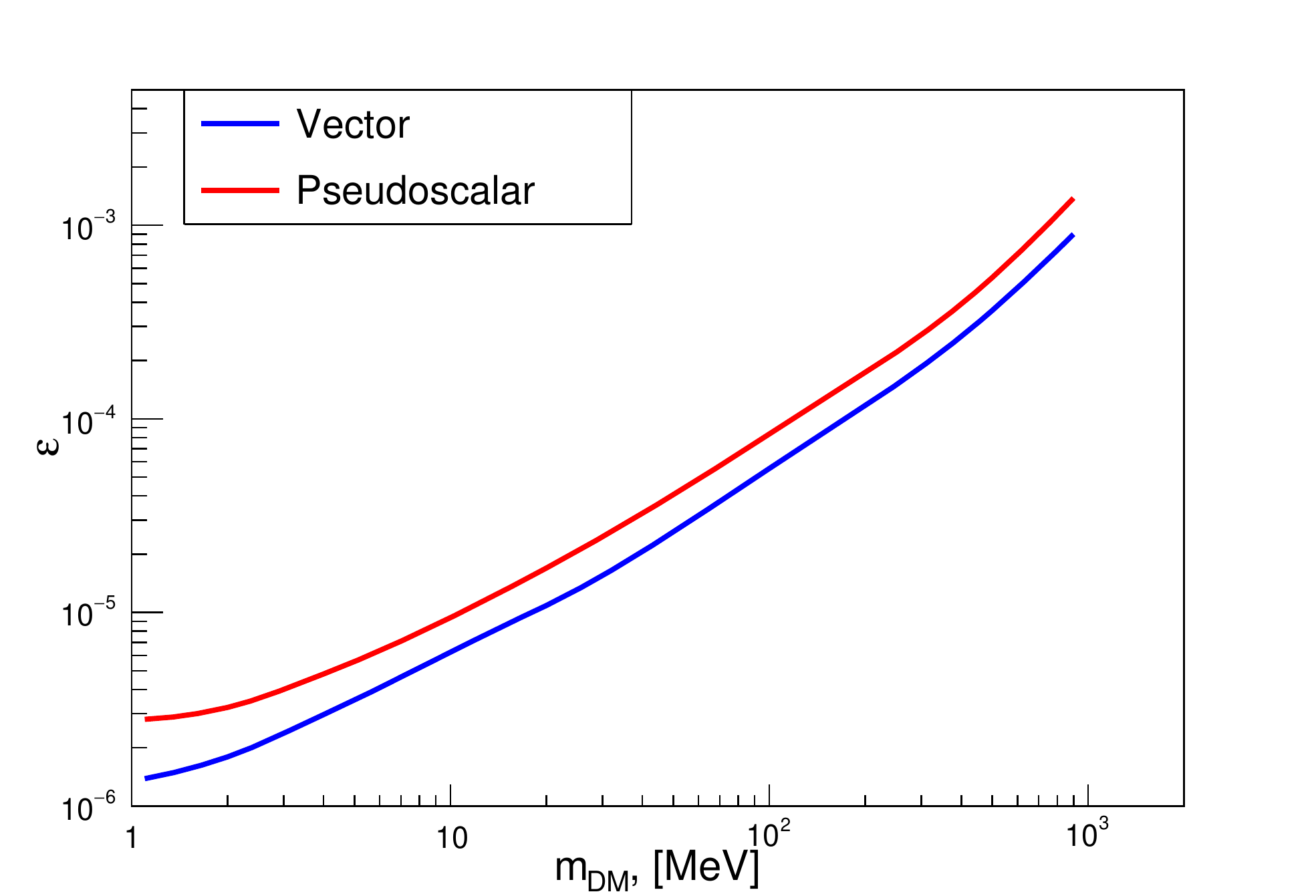}
\caption {Sensitivity of the missing energy signature experiment to vector and pseudoscalar DM for $5\times10^{12}$ EOT.
\label{fig:sensitivity1}}
\end{center}
\end{figure*}

\begin{figure*}[ht]
\centering
\includegraphics[width=\linewidth]{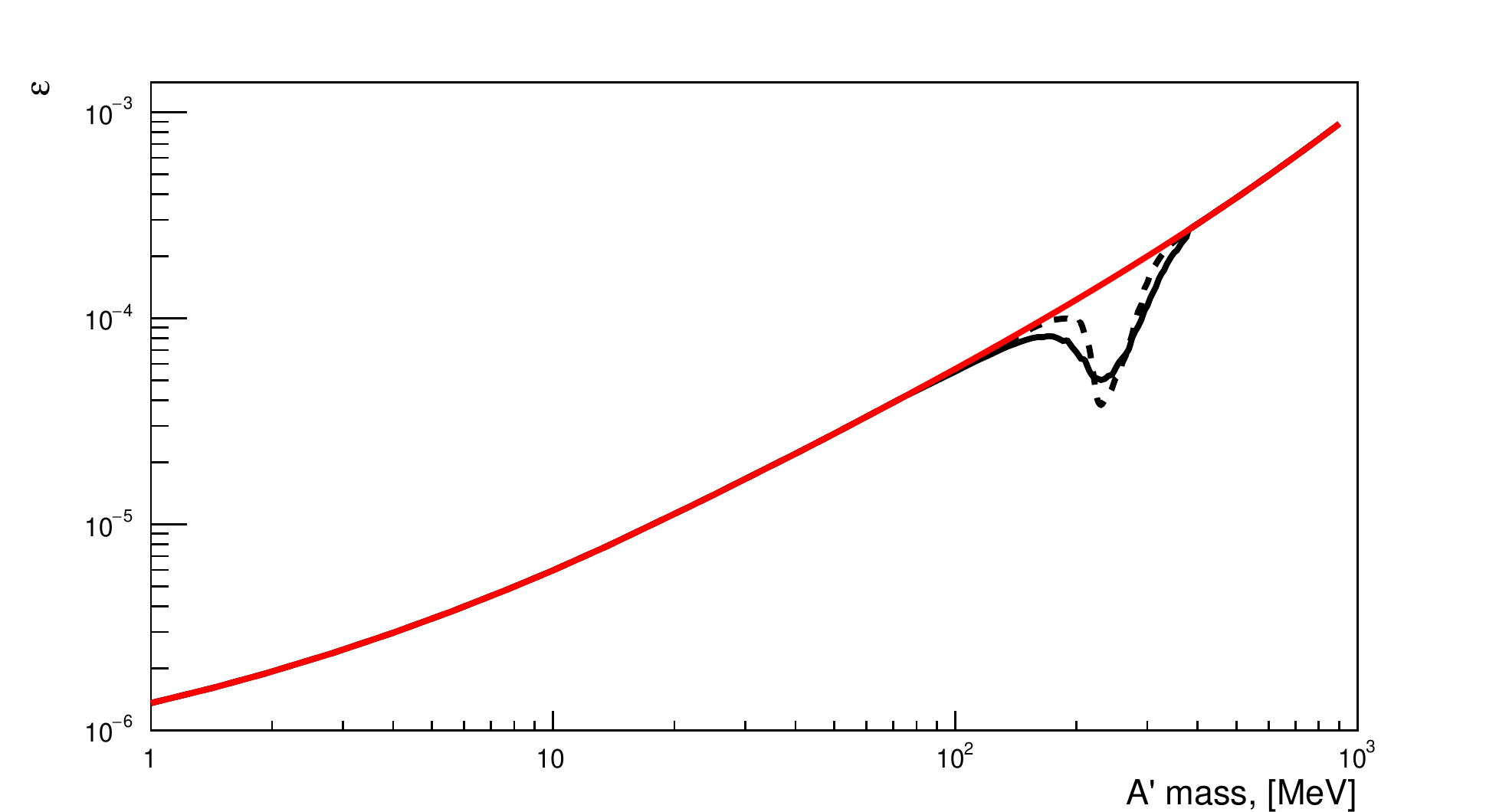}
\caption{Sensitivity of the missing energy signature experiment to vector DM for $5\times10^{12}$ EOT.
          The sensitivity that takes into account the contribution from the annihilation process for $\alpha_D=0.5$ ($\alpha_D=0.1)$
          is shown by the black continuous (dashed) line.}
\label{fig:sensitivity2}
\end{figure*}

 We calculated the sensitivity of the missing energy signature fixed target experiment to light DM particles
for the statistics corresponding to $5\times10^{12}$ EOT assuming the background-free conditions and 100\% efficiency
of the experiment. The result for the vector and pseudoscalar mediators, with only bremsstrahlung processes taken into
account, is shown in Figure~\ref{fig:sensitivity1}. The contribution from the annihilation processes is significant
at the masses above 100 MeV, but it is more model-dependent. The corresponding sensitivity for the two values of $\alpha_D$
is shown in Figure~\ref{fig:sensitivity2}.

\section{Conclusion}

 The package DMG4 for the simulation of light dark matter production in fixed target experiments is created. It can be
used in simulation programs of experimental setups based on the Geant4 framework. As an example, we calculated the sensitivity
of a typical missing energy signature experiment to various types of light dark matter.

 The package is available at http://mkirsano.web.cern.ch/mkirsano/DMG4.tar.gz. It is recommended also to contact the corresponding
author Mikhail Kirsanov about the usage.

\section{Aknowledgements}

  This work was supported by the Ministry of Science and Higher Education (MSHE) and RAS (Russia), Tomsk Polytechnic University within
the assignment of MSHE (Russia), the European Research Council (ERC) under the European Union’s Horizon 2020 research and innovation
program (Grant agreement No. 947715 - POKER Starting Grant).

\section{Appendix A}

In this section we collect brems-like differential cross-sections
of the processes $l N \to l N X$, where $X=(S,P,V,A)$ and 
$l = (e^\pm, \mu^\pm)$. For the IWW approach~\cite{Liu_2017,Liu_2017_2}
one has the following expressions for the cross-sections
\begin{equation}
\left(\frac{d \sigma^X}{dx \, d\cos \theta} \right)_{IWW}=
2 \epsilon_X^2 \alpha^3 | {\bf k}| E_0 (1-x) \frac{\chi}{\tilde{u}^2} 
|\mathcal{A}^X|^2
\end{equation}
where $x=E_X/E_0$ is the energy fraction that DM mediators carry away, 
$\theta$ is the emission angle of DM mediators, $|{\bf k}| = \sqrt{E_X^2-m_X^2}$ is the momentum of hidden $X$-bosons, $E_0$ is the initial 
energy of the incident particle in the beam, 
$\tilde{u} = -x E_0^2 \theta^2 - m_X^2(1-x)/x-m_l^2 x$ is the approximate value for the auxiliary Mandelstam variable,
$\chi$ is the standard photon flux that takes into account the elastic form-factors $F_{el}(t)$. The corresponding expressions for $\chi$ and $F_{el}(t)$
can be found elsewhere~\cite{DMsimulation}. The expressions for amplitudes squared are~\cite{Liu_2017,Liu_2017_2}
\begin{align}\label{ListOfAmpSq}
|\mathcal{A}^S|^2 = & \frac{x^2}{1-x}+2(m_S^2-4m_l^2)\frac{\tilde{u}x+m_S^2(1-x)+m_l^2x^2}{\tilde{u}^2},\nonumber\\
|\mathcal{A}^{P}|^2 = &\frac{x^2}{1-x}+2m_P^2\frac{\tilde{u}x+m_P^2(1-x)+m_l^2x^2}{\tilde{u}^2},\nonumber\\
|\mathcal{A}^{V}|^2 = &2\frac{2-2x+x^2}{1-x}+4(m_V^2+2m_l^2)\frac{\tilde{u}x+m_V^2(1-x)+m_l^2x^2}{\tilde{u}^2}\\
|\mathcal{A}^{A}|^2 = &\frac{4m_l^2x^2}{m_A^2(1-x)}+2\frac{2-2x+x^2}{1-x}+4(m_A^2-4m_l^2)\frac{\tilde{u}x+m_A^2(1-x)+m_l^2x^2}{\tilde{u}^2}\nonumber.
\end{align}

\section{Appendix B}

In this section we place various figures referenced in the main sections of the paper.

\clearpage
\bibliography{bibliographyOther,bibliographyNA64,bibliographyNA64exp}

\begin{thebibliography}{10}
\expandafter\ifx\csname urlstyle\endcsname\relax
  \providecommand{\doi}[1]{doi:\discretionary{}{}{}#1}\else
  \providecommand{\doi}{doi:\discretionary{}{}{}\begingroup
  \urlstyle{rm}\Url}\fi

\bibitem{HOLDOM1986196}
B.~Holdom.
\newblock \emph{Phys. Lett. B} 166, 196  (1986).
\newblock \doi{10.1016/0370-2693(86)91377-8}.

\bibitem{mb}
M.~Battaglieri, et~al.
\newblock {US} cosmic visions: New ideas in dark matter 2017: Community report
  (2017).
\newblock ArXiv:1707.04591 [hep-ph].

\bibitem{Beacham:2019nyx}
J.~Beacham et~al.
\newblock \emph{J. Phys. G} 47, 010501 (2020).
\newblock \doi{10.1088/1361-6471/ab4cd2}.

\bibitem{jdb}
J.~D. Bjorken, et~al.
\newblock \emph{Phys. Rev. D} 80, 075018 (2009).
\newblock \doi{10.1103/PhysRevD.80.075018}.

\bibitem{PhysRevD.88.114015}
E.~Izaguirre, et~al.
\newblock \emph{Phys. Rev. D} 88, 114015 (2013).
\newblock \doi{10.1103/PhysRevD.88.114015}.

\bibitem{PhysRevD.80.095024}
B.~Batell, M.~Pospelov, and A.~Ritz.
\newblock \emph{Phys. Rev. D} 80, 095024 (2009).
\newblock \doi{10.1103/PhysRevD.80.095024}.

\bibitem{PhysRevD.91.094026}
E.~Izaguirre, et~al.
\newblock \emph{Phys. Rev. D} 91, 094026 (2015).
\newblock \doi{10.1103/PhysRevD.91.094026}.

\bibitem{geant4}
S.~Agostinelli et~al.
\newblock \emph{Nucl. Instrum. Meth. A} 506, 250 (2003).
\newblock \doi{10.1016/S0168-9002(03)01368-8}.

\bibitem{PDG_particles}
{Monte Carlo Particle Numbering Scheme}.
\newblock
  \url{https://pdg.lbl.gov/2019/reviews/rpp2019-rev-monte-carlo-numbering.pdf}.
\newblock Accessed: 2021-01-25.

\bibitem{dusaev2020photoproduction}
R.~R. Dusaev, D.~V. Kirpichnikov, and M.~M. Kirsanov.
\newblock \emph{Phys. Rev. D} 102, 055018 (2020).
\newblock \doi{10.1103/PhysRevD.102.055018}.

\bibitem{PhysRevLett.121.041802}
L.~Marsicano, et~al.
\newblock \emph{Phys. Rev. Lett.} 121, 041802 (2018).
\newblock \doi{10.1103/PhysRevLett.121.041802}.

\bibitem{Liu_2017}
Y.-S. Liu, D.~McKeen, and G.~A. Miller.
\newblock \emph{Physical Review D} 95, 036010 (2017).
\newblock \doi{10.1103/physrevd.95.036010}.

\bibitem{Liu_2017_2}
Y.-S. Liu and G.~A. Miller.
\newblock \emph{Physical Review D} 96, 016004 (2017).
\newblock \doi{10.1103/physrevd.96.016004}.

\bibitem{Dobrich_2016}
B.~Döbrich, et~al.
\newblock \emph{Journal of High Energy Physics} 2016 (2016).
\newblock \doi{10.1007/jhep02(2016)018}.

\bibitem{DMsimulation}
S.~Gninenko, et~al.
\newblock \emph{Phys. Lett. B} 782, 406  (2018).
\newblock \doi{10.1016/j.physletb.2018.05.010}.

\bibitem{Mathematica}
W.~R. Inc.
\newblock Mathematica, {V}ersion 12.2.
\newblock Champaign, IL, 2020.

\bibitem{na64-prd}
D.~Banerjee, et~al.
\newblock \emph{Phys. Rev. D} 97, 072002 (2018).
\newblock \doi{10.1103/PhysRevD.97.072002}.

\end{thebibliography}
\bibliographystyle{na64-epjc}



    


\end{document}